# Conversion of a force curve between chemically the same surfaces into the number density distribution of the particles on the surface using a structure factor

**Ken-ichi Amano, Kota Hashimoto, Ryosuke Sawazumi**

*Department of Energy and Hydrocarbon Chemistry, Graduate School of Engineering, Kyoto University, Kyoto 615-8510, Japan.*

Author to whom correspondence should be addressed: Ken-ichi Amano.
Electric mail: amano.kenichi.8s@kyoto-u.ac.jp

**ABSTRACT:** Line optical tweezer and colloidal-probe atomic force microscopy can measure force curves between two large colloidal particles of chemically the same surfaces in a suspension of small colloidal particles. Recently, the authors proposed a transform theory to obtain the number density distribution of the small colloidal particles on the large colloidal particle from the force curve. In this short letter, we propose another method which utilizes Ornstein-Zernike equation coupled with a closure equation instead of Kirkwood superposition approximation. The new transform theory uses a structure factor measured by x-ray or neutron scattering, and applies Nelder-Mead method to find the solution. Since it is known that Ornstein-Zernike equation coupled with the closure equation is accurate compared with Kirkwood superposition approximation, the new transform theory is theoretically better than the previous methods when the structure factor and the closure equation are reliable.

**MAIN TEXT**

It is known that line optical tweezer (LT) [1] and colloidal-probe atomic force microscopy (CP-AFM) [2,3] can measure a force curve between two large colloidal particles of chemically the same surfaces in a suspension of small colloidal particles. For instance, the force curve ($f$) is measured in order to estimate the number density distribution of the small colloidal particles on the large colloidal particle ($\rho$). However, $\rho$ cannot be directly obtained from LT and CP-AFM. Hence, recently, Amano proposed several transform theories that can transform from $f$ into $\rho$. The first transform theory [4] is proposed by applying one-dimensional Ornstein-Zernike equation coupled with a closure equation (1D-OZ-closure). It uses the total correlation function between the small colloidal particles ($h_{SS}$) and a trial function of the total correlation function between the large and small colloidal particles ($h_{LS}$). Since the author was unfamiliar with the convergence technique, the first transform theory could not reach a sufficiently proper solution and the theory itself vanished in smoke. Secondly, the author proposed a transform theory that uses Kirkwood superposition approximation [5–8]. It can provide semi-quantitatively reliable results if the input $f$ is reliable, but interlayer compression induced by sandwiching the suspension between the two large colloidal particles could not be corrected due to the Kirkwood superposition approximation. In order to obtain more accurate $\rho$, we should use a modified Kirkwood superposition approximation or propose another method without it. In the short letter, we retry to use 1D-OZ-closure again to overcome the problem. For the calculation, a structure factor measured by x-ray or neutron scattering is used as an input, and Nelder-Mead (NM) method [9] is used in the convergence process.

To obtain $\rho$, firstly, we extract a force curve ($f_S$) induced by small colloidal particles from $f$ by using equation below:

$$f_S = f - f_0, \qquad (1)$$

where $f_0$ represents the force curve between the large spheres in a solution where there are no small colloidal particles. We prepare the structure factor ($S(q)$) in the (bulk) suspension of the small colloidal particles, where $q$ is length of the scattering vector. $S(q)$ is changed to the total correlation function between the small colloidal particles in reciprocal space ($H_{SS}(q)$) by using a following equation:

$$H_{SS}(q) = [S(q) - 1]/\rho_0, \qquad (2)$$

where $\rho_0$ represents the bulk number density of the small colloidal particles. Outline of the convergence process is as follows:

(I) Before starting NM method, prepare the total correlation function between the large and small colloidal particles ($h_{LS}$). The shape of $h_{LS}$ is arbitrary, which is generated randomly or empirically (the values of $h_{LS}$ in the overlap region are set at $-1$). The function $h_{LS}$ is a discrete function, and the segmentation number is $N$. For NM method performed in step (VI), the number of $h_{LS}$ is $N+1$. We notify that all of the shapes of $h_{LS}$ must be different each other.

(II) Calculate the indirect correlation function between the large colloidal particles ($w_{LL}$) by using 1D-OZ:

$$w_{LL} = \text{FT}^{-1}[\rho_0 H_{LS}(q)^2/(1 + \rho_0 H_{SS}(q))], \qquad (3)$$

where $\text{FT}^{-1}$ represents the inverse Fourier transform, and $H_{LS}(q)$ is the total correlation function between the large and small colloidal particles in the reciprocal space. All of the $h_{LS}$ prepared in step (I) are changed to each $w_{LL}$ through Eq. (3).

(III) Obtain potential between the large colloidal particles solely induced by the small colloidal particles ($\varphi$) using a closure equation. For example, when the closure equation is the hypernetted-chain (HNC) approximation, $\varphi$ is equal to $k_B T w_{LL}$, where $k_B$ and $T$ represent the Boltzmann constant and the absolute temperature, respectively. When a bridge function ($b$) is incorporated into HNC approximation, $\varphi$ is equal to $k_BT(w_{LL} + b)$. When the closure equation is Percus-Yevick (PY) approximation, $\varphi$ can be calculated as $-k_BT\ln(1 + w_{LL})$. All of the functions of $w_{LL}$ obtained in step (II) are changed to each $\varphi$. We notify that the closure equation used in this step should be selected according to the experimental condition.

(IV) Differentiate $-\varphi$ with respect to the distance between the large colloidal particles, and obtain a force curve between the large colloidal particles induced by the small colloidal particles being $f_{Scal}$. Then, there are $N + 1$ numbers of the $f_{Scal}$ on this stage.

(V) Prepare the error value being $E$ from $f_{Scal}$ by using Eqs. (4)-(6):

$$E = E_1 + E_2, \quad (4)$$

$$E_1 = \int_R^\infty \alpha[f_{Scal}(r) - f_S(r)]^2 dr, \quad (5)$$

$$E_2 = \varepsilon P[f_S(r), f_{Scal}(r)], \quad (6)$$

where $E_1$ is an error sum of squares, $E_2$ is an error value originates from a penalty function $P$ composed of $f_S$ and $f_{Scal}$, $r$ is distance between the large colloidal particles, $\alpha$ is $r^2$, $r$, or 1, and $\varepsilon$ is a positive weight coefficient. A handling process of the pair of $\alpha$ and $\varepsilon$ is written in the next step. Experimentally, the large colloidal particles cannot be approached only up to a certain distance, and then $R$ corresponds to the distance. The penalty function is introduced to elaborate the smooth fitting of $f_{Scal}$ on $f_S$. Moreover, the penalty function can be applied to smooth the shape of $f_{Scal}$ in the region less than $R$. The number of $E$ is $N + 1$.

(VI) Perform NM method and find an optimized $h_{LS}$. Since it is difficult to obtain a

sufficiently optimized $h_{LS}$ from one time NM method, repeat NM method around the newly obtained $h_{LS}$ until the error value $E$ becomes sufficiently small. (We recommend that coefficients $\alpha$ and $\varepsilon$ are changed at each repeat, and $\varepsilon$ is set at 0 at the final NM method.) Consequently, one optimized function $h_{LS}$ is obtained.

(VII) Perform above steps by using the other inputs, and obtain another $h_{LS}$. Then, calculate averaged $h_{LS}$ (this is the answer we would like to obtain). If the obtained $h_{LS}$ is proper, the function does not have unnaturally short-period oscillations. The values of $h_{LS}$ in all the regions should be greater than or equal to $-1$, and $h_{LS}$ should start from $-1$ and end with 0.

(VIII) Calculate $H_{LS}(q)$ from $h_{LS}$. Using 1D-OZ, $H_{LS}(q)$, and $H_{SS}(q)$, obtain the direct correlation function between the large and small colloidal particles ($c_{LS}$). Then, two-body potential between the large and small colloidal particles ($u_{LS}$) is estimated by using the closure equation, $h_{LS}$, and $c_{LS}$. (The function $u_{LS}$ is also one of the answers we would like to obtain.)

In summary, we have proposed the new transform theory which uses 1D-OZ-closure instead of Kirkwood superposition approximation. It is expected that the new transform theory can reproduce more accurate $h_{LS}$ compared with the previous one, when the input data of $S(q)$ and the closure equation are reliable. In the future, we will conduct a verification test of the new transform theory within a computationally closed cycle.

**ACKNOWLEDGEMENTS**

We thank T. Sakka and N. Nishi for helpful discussion. This work was supported by "Grant-in-Aid for Young Scientists (B) from Japan Society for the Promotion of Science (15K21100)".